\documentclass[superscriptaddress,showpacs,a4paper,article,twocolumns]{revtex4}

\usepackage{graphicx}
\usepackage[latin1]{inputenc}

\begin{document}

\title{Faraday optical isolator in the 9.2~$\mu$m range for QCL applications}

\author{Laurent Hilico}
\affiliation{D\'epartement de Physique, Universit\'e d'Evry Val d'Essonne, Boulevard F. Mitterrand, 91025 Evry cedex}
\affiliation{Laboratoire Kastler Brossel, UPMC-Paris 6, ENS, CNRS ; Case 74, 4 place Jussieu, 75005 Paris, France}
\author{Albane Douillet}
\affiliation{D\'epartement de Physique, Universit\'e d'Evry Val d'Essonne, Boulevard F. Mitterrand, 91025 Evry cedex}
\affiliation{Laboratoire Kastler
Brossel, UPMC-Paris 6, ENS, CNRS ; Case 74, 4 place Jussieu, 75005 Paris, France}
\author{Jean-Philippe Karr}
\affiliation{D\'epartement de Physique, Universit\'e d'Evry Val d'Essonne, Boulevard F. Mitterrand, 91025 Evry cedex}
\affiliation{Laboratoire Kastler Brossel, UPMC-Paris 6, ENS, CNRS ; Case 74, 4 place Jussieu, 75005 Paris, France}
\email{hilico@spectro.jussieu.fr}
\author{Eric Tourni\'e}
\affiliation{Universit\'e Montpellier 2, Institut d'Electronique du Sud-UMR 5214 CNRS, Place Eug\`ene Bataillon, 34095 Montpellier Cedex 5, France.}

\begin{abstract}
We have fabricated and characterized a n-doped InSb Faraday isolator in the mid-IR range (9.2~$\mu$m). A high isolation ratio of $\approx$30~dB with a transmission over 80\% (polarizer losses not included) is obtained at room temperature. Further possible improvements are discussed. A similar design can be used to cover a wide wavelength range ($\lambda \sim 7.5-30\ \mu$m).
\end{abstract}

% \ocis{230.3240, 160.3820, 140.5965, 140.3070, 020.4180.}
\pacs{ocisnumbers 230.3240, 160.3820, 140.5965, 140.3070, 020.4180}
\maketitle

\section{Introduction}

Recent progress in quantum cascade laser (QCL) technology towards high power cw operation and off-the-shelf availability for a widening wavelength range has enlarged their application fields, for instance in molecular spectroscopy, lidar applications or THz imaging. Thanks to their wide tunability and integrability, QCLs are superseding alternative sources such as molecular gas lasers or lead-salt lasers~\cite{faist94,capasso10,hugi10}. One of the main drawbacks of QCLs is their extreme sensitivity to optical feedback that may prevent stable operation, hence the need for high optical isolation, especially for high-resolution spectroscopy applications in which precise frequency control is required.

QCLs are linearly polarized sources. A simple way to achieve optical isolation is to use a wire grid polarizer and a quarter-wave plate~\cite{mansfield80}. However, the price to pay is rather high since the laser beam has to keep a circular polarization: any modification of the feedback beam polarization reduces the isolation ratio. The alternative solution relies on the Faraday effect, and was studied between the 60's and the 90's by several authors in the context of CO$_2$ laser applications. Mid-infrared Faraday isolation has been demonstrated using several Faraday media and different processes such as free carrier contribution at room temperature or interband and free carrier spin contributions at cryogenic temperatures \cite{dennis68,boord74,jacobs74,tomasetta79,aggarwal88,carlisle89,klein89}. Isolation ratios up to 30~dB and insertion losses of 1.5~dB (71~\% transmission) have been reported. n-doped InSb, which benefits from a strong free carrier Faraday effect, appears to be the most promising material in the 9-10~$\mu$m range~\cite{boord74}. Nevertheless, it also suffers from rather large optical absorption and low heat conductivity, which make it unsuited for isolation of high-power CO$_2$ lasers. For this reason, development efforts were stopped, and currently no mid-IR Faraday isolator is commercially available.

Our initial motivation to fabricate such a device is the experiment developed in the Paris group, which we briefly describe here to illustrate an application requiring both a high isolation ratio and linearly polarized light. The experiment aims at a high-precision measurement of a Doppler-free two-photon vibrational transition frequency in the H$_2^+$ molecular ion in the 9.2~$\mu$m range in order to obtain a new determination of the proton-to-electron mass ratio~\cite{roth08,korobov09}. We developed a cw QCL source with a frequency control at the kHz level by a phase lock against a stabilized CO$_2$ laser~\cite{bielsa07,bielsa08}. The QCL beam ($\approx$54~mW) is mode-matched to a high finesse~($\approx$1000) Fabry-Perot cavity, that is required both to ensure a Doppler-free geometry with perfectly counterpropagating beams, and to enhance the transition rate (two-photon ro-vibrational transitions in H$_2^+$ being very weak). Optical feedback from the high-finesse cavity is
  a problem in such an experiment; in the initial setup, isolation was obtained by combining a quarter-wave plate and polarizer with an acousto-optic modulator providing a 6~dB additional isolation through its strongly polarization-dependent efficiency~\cite{karr08a}. The total isolation ratio, slightly over 30~dB, was just sufficient to achieve stable operation of the QCL. However, probing the transitions with a circularly polarized beam has two important drawbacks: (i) two-photon transition rates in H$_2^+$  are almost an order of magnitude smaller with respect to the linear polarization case~\cite{karr08a}, and (ii) transition frequencies are much more sensitive (by 4-5 orders of magnitude) to the Zeeman shift due to the $|\Delta M| = 2$ selection rule~\cite{karr08b}, requiring a high level of magnetic field control.

In this paper, we report the performances of the device we have fabricated, which is based on the same material (n-doped InSb) as most earlier realizations, but takes advantage of improvements in wafer quality and permanent magnets. After a brief review of Faraday effect in InSb, we describe the isolator setup and our experimental results. In the last section, we compare the performances of our device with published data and discuss the feasibility of a double-stage mid-infrared Faraday isolator.

\section{Free carrier Faraday effect in InSb}

At room temperature, infrared absorption and Faraday effect in n-doped InSb are dominated by the free electron contribution~\cite{boord74} and can be described in the frame of the Drude model. The polarization rotation in a magnetic field B for a wafer thickness L is given by $\theta=VBL$ where
\begin{equation}
V=\frac{\theta}{BL}=\frac{\mu_0 N q^3\lambda^2}{8\pi^2\ n\ m^{*2}\ c}\label{eq_exp_verdet},
\end{equation}
is the Verdet constant~\cite{Boer90}, and the absorption coefficient $\alpha$ is
\begin{equation}
\alpha=\frac{\mu_0\ N\ q^2\ \lambda^2}{4\pi^2\ n\ c\ m^*\ \tau}\label{alpha},
\end{equation}
where $n=4$ is the material refractive index, $m^*$ is the carrier effective mass, $N$ the carrier density and $\tau$ the effective carrier relaxation time. Equation~(\ref{eq_exp_verdet}) (resp.~(\ref{alpha})) are valid if the conditions $\omega_r\ll\omega_c\ll\omega\leq\omega_p\ll n\omega$ (resp. $\omega_r\ll\omega$) are satisfied, where $\omega_c$ is the cyclotron frequency, $\omega$ the optical field frequency, $\omega_p$ the plasma frequency, and $\omega_r = 1/\tau$. Note that both Faraday rotation and absorption follow a $\lambda^2$ law.

The relevant figure of merit when assessing the performances of a material for optical isolation is the Faraday rotation $F_m$ per dB of attenuation and per Tesla. When assessing device performances (with a given magnetic field), it is best to use the Faraday rotation per dB attenuation $F_d$. They are given by
\begin{eqnarray}\label{F}
    F_m&=&\frac{\theta}{\alpha L B}=\frac{qc\tau}{2\mu_0\ m^*}\\
    F_d&=&F_m B.
\end{eqnarray}
In order to optimize $F_d$, one should have a $B$-field as strong as possible, and semiconductors of low effective mass and low resistivity, as discussed in~\cite{boord74}. The figures of merit appear to be independent of the carrier density $N$, but a more detailed approach allows to study the influence of donor concentration and temperature~\cite{boord74}; for InSb at 300 K, the optimal value is around $N \approx$~2.10~$^{17}$~cm$^{-3}$. We measured $N = 2.35$ $10^{17}$ to $2.5$ $10^{17}$~cm$^{-3}$ in our samples (see Table~\ref{tab_wafer_charact}) by Hall effect measurements.

Let us check that the validity condition of equations~(\ref{eq_exp_verdet}-\ref{alpha}) is met in our experimental conditions. The carrier effective mass in InSb is $m^*\approx$~0.014$\ m_e$~\cite{ioffe-web}; taking as carrier density $N = 2.35$ $10^{17}$~cm$^{-3}$, the plasma frequency is $\omega_p=(\mu_0 N q^2 c^2/m^*)^{1/2} \approx2\pi \times$37~THz. With B~$\approx$~1~T the cyclotron frequency is $\omega_c\approx2\pi \times2$~THz. The effective carrier relaxation time $\tau$ is $\approx$~1.4~10$^{-12}$~s, so $\omega_r \approx2\pi \times0.1$~THz and the laser frequency is $\omega\approx2\pi \times32$~THz.
One has indeed $\omega_r\ll\omega_c\ll\omega\leq\omega_p\ll n\omega$.

\section{Experimental setup}

\begin{figure}
\includegraphics[width=7cm]{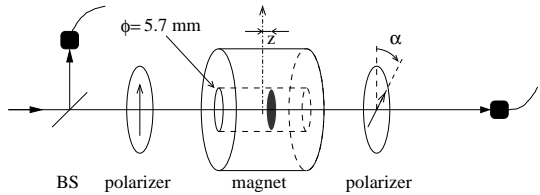}\label{fig_manip}
\caption{Experimental setup. $z$ is the wafer position with respect to the magnet center. The black boxes are power meters.}
\end{figure}

\begin{figure}
\includegraphics[width=7cm]{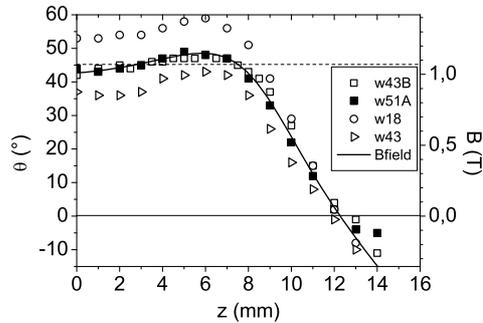}\label{fig_rotation}
\caption{Symbols: Faraday rotation in deg versus wafer position z in the magnet at $\lambda=$~9.166~$\mu$m. The sample thickness is 390$\mu$m except for w18 which is 520~$\mu$m-thick. Solid line: magnetic field profile. The dashed line indicates 45~deg rotation.}
\end{figure}

The experimental setup is depicted in figure~\ref{fig_manip}. The longitudinal magnetic field is produced by a standard permanent
magnet. The solid line in figure~\ref{fig_rotation} shows the $B$-field profile. The maximum field is 1.16~T. Four high-quality Te n-doped InSb wafers
from Wafer Technology were cut into 4x4 mm square plates, antireflection coated and glued on a 0.5~mm pitch hollow copper screw to adjust the
wafer position in the magnet. The wafers come from different slices of the same InSb ingot and exhibit slightly different
characteristics that are reported in Table~\ref{tab_wafer_charact}. They are labeled 18, 43, 43B and 51 A.

All intensity measurements were done using two power-meters, one being used as a reference to cancel out possible laser power drifts. The wafer reflection coefficient was measured at small incidence to separate the incident and reflected beams. The residual reflectivity is $\leq$~3\%, instead of 36\% without coating. The transmission coefficient $T$ at $\lambda = 9.166 \mu$~m (measured without polarizers) lies between 74 and 80\% at room temperature for all samples. For Faraday rotation measurements, one wire grid polarizer was installed at each end of the Faraday rotator. The polarization rotation was monitored by rotating the second polarizer so as to minimize the transmitted intensity.

\section{Results}

\begin{figure}
\includegraphics[width=7cm]{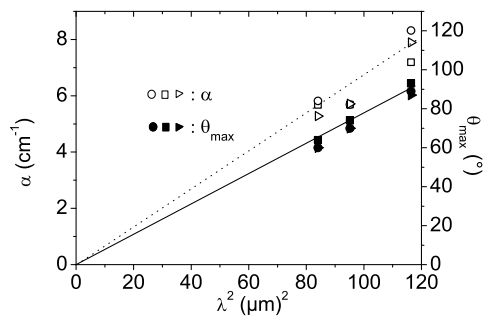}\label{fig_absorption_rotation}
\caption{Maximum Faraday rotation (filled symbols) and absorption coefficient (open symbols) for three 520~$\mu$m-thick samples, versus the wavelength squared. The lines are adjustments by a proportional law with slopes of 0.76~deg($\mu$m)$^{-2}$ and 0.065~cm$^{-1}$($\mu$m)$^{-2}$, respectively.}
\end{figure}

First, we measured the Faraday rotation angles and absorption coefficients of three 520~$\mu$m-thick samples at three different wavelengths, using a CO$_2$ laser oscillating on the 9R42 (9.166~$\mu$m), 9P42 (9.753~$\mu$m) and 10P38 (10.787~$\mu$m) lines. Figure~\ref{fig_absorption_rotation} shows that both follow a $\lambda^2$ law, with respective slopes of 0.0065~cm$^{-1}$($\mu$m)$^{-2}$ and 0.76~deg($\mu$m)$^{-2}$. Comparison with equations ~(\ref{eq_exp_verdet}-\ref{alpha}) yields $m^*$=0.026~$m_e$ and $\tau$=1.4~10$^{-12}$~s. This effective mass is larger than the value of~\cite{ioffe-web}, but consistent with the value of 0.022 given in~\cite{Zawadski}.

Since polarization rotations of about 60~deg were measured at 9.166~$\mu$m, the wafers were thinned to 390~$\mu$m to obtain rotation by about 45~deg for the maximum $B$-field. Figure~\ref{fig_rotation} shows the polarization rotation as a function of the wafer position $z$ in the magnet. As expected, it follows the $B$-field profile, and the value of the Verdet constant (see eq.~(\ref{eq_exp_verdet})) can be extracted from this data. All characteristics are reported in Table~\ref{tab_wafer_charact}, as well as the corresponding figure of merit, showing that wafer 51A has the best performances.

\begin{table}[h]
\center
\begin{tabular}{|c|c|c|c|c|c|c|c|c|}
  \hline
        & $L$    & $N$            & $\theta_{max}$  & $V$                    &  $R$  &  $T$  & $\alpha$  & $F_d$ \\
  wafer & $\mu$m & cm$^{-3}$      &     deg.        & deg. T$^{-1}$mm$^{-1}$ &  \%   &  \%   & cm$^{-1}$ & deg.T$^{-1}$\\
  \hline
  43    & 390    & 2.5  10$^{17}$ & 44              & 97.3                   & 3.1   & 78.4  & 4.63      & 210\\
  18    & 520    & 2.5  10$^{17}$ & 59              & 97.8                   & 3.1   & 74.0  & 4.6       & 213\\
  43B   & 390    & 2.35 10$^{17}$ & 48              & 106                    & 2.8   & 76.1  & 5.54      & 191\\
  51A   & 390    & 2.35 10$^{17}$ & 49              & 108                    & 3.7   & 79.8  & 3.86      & 280\\
  \hline
\end{tabular}
\caption{Wafer characteristics: thickness $L$, carrier density $N$, polarization rotation maximum angle $\theta_{max}$, Verdet constant $V$, reflection and transmission coefficients $R$ and $T$, absorption coefficient $\alpha$ and wafer figure of merit $F_d$, for $\lambda$=9.166~$\mu$m.} \label{tab_wafer_charact}
\end{table}

The minimum transmitted intensity observed in polarization rotation measurements gives the Faraday isolation ratio. We measured between 27~dB and 30~dB, a value limited by the polarizer extinction ratio and by the power meter accuracy, but not by $B$-field or wafer dopant concentration
inhomogeneities. The measured wire grid polarizer transmission is 90~\%, so the overall transmission of the isolator
is $0.8\times 0.9^2=65~\%$ corresponding to 1.9~dB insertion loss.

Since Faraday rotation decreases and absorption increases with temperature, a simple way to improve the performances consists in lowering the wafer temperature~\cite{boord74}. The isolator temperature was varied using Peltier coolers. Table~\ref{tab_temperature_effect} give the maximum Faraday rotation and wafer transmission at 14, 24 and 34 $^{\circ}$C for the best wafer (51A). Wafer transmissions over 80\% are achieved. The temperature dependance of the Faraday rotation angle is 1.5~deg~K$^{-1}$. This also shows that it is necessary to stabilize the isolator temperature to better than 1~K to ensure long-term stability of the isolation ratio in the 30~dB range.

\begin{table}[h]
\center
\begin{tabular}{|c|l|c|}
  \hline
  $T_0$ ($^{\circ}$C) & $\theta_{max}$ & $T$~(\%) \\
  \hline
  14 & 50 & 80.4 \\
  24 & 48.5 & 78.5 \\
  34 & 47 & 76.8 \\
  \hline
\end{tabular}
\caption{Maximum polarization rotation and 51A wafer transmission versus the temperature $T_0$.}\label{tab_temperature_effect}
\end{table}

We checked that this optical isolator (operated at 16$^{\circ}$C to avoid water condensation) could be used for intracavity two-photon spectroscopy. The QCL beam was injected in the high finesse Fabry-Perot cavity (see Introduction), and locked on resonance. By carefully adjusting the polarizers, we obtained stable operation but only over a few minutes, indicating that the isolation ratio is only barely sufficient for this application.

As discussed in the Introduction, this result implies a significant increase of the transition probabilities in H$_2^+$ two-photon spectroscopy. For precise comparison with the formerly used polarizer/quarter-wave plate system, insertion losses have to be taken into account. When the Faraday isolator is used, an additional half-wave plate of transmission 98~\% is required to adjust the polarization, so that the overall transmission is $T = 0.8$ (wafer) $\times 0.9^2$ (polarizers) $\times 0.98 = 0.635$. The transmission of the former system is $0.9$ (polarizer) $\times 0.98$ (quarter-wave plate) $= 0.88$. Since the intensity of the considered two-photon lines is 8.5 larger in linear polarization~\cite{karr08a}, and the transition rate being proportional to the square of the laser intensity, the net gain is a factor $8.5 \times (0.635/0.88)^2 = 4.4$.

\section{Discussion}

Table~\ref{tab_Faraday_comp} summarizes published characteristics of Faraday isolators in the 9-10~$\mu$m range and at 5.4~$\mu$m together with the present results. Note that the measurements reported in~\cite{dennis68} were performed without antireflection coating on the wafer, leading to inaccurate absorption coefficient and figures of merit. The cryogenic temperature isolator of Ref.~\cite{tomasetta79} clearly exhibit the best figures of merit, but a rather low isolation ratio, with the drawback of a complicated setup. Due to the high $B$-field and good wafer quality, our isolator shows improved performance as compared to previously published ones at room temperature.

\begin{table}[h]
\center \tiny
\begin{tabular}{|c|c|c|r|l|c|l|c|c|c|c|}
\hline
Material &    $N$       & $\lambda$ & $T_0$ & \multicolumn{1}{c|}{$B$}  & isolation & \multicolumn{1}{c|}{$\alpha$}    & Wafer ins. & $F_m$          &
  $F_d$      & \\
         & (cm$^{-3}$)  & ($\mu$m)  & (K)   & \multicolumn{1}{c|}{(T)}  &   (dB)    & \multicolumn{1}{c|}{(cm$^{-1}$)} &  loss (dB) & (deg T$^{-1}$) &
  (deg)      & ref.\\
  \hline
  InSb & 2.0\ 10$^{17}$  & 10.6  & 300 & 0.53     & 30 & 2.1   & 0.5 & 809  & 429  & \cite{dennis68}\\
  InSb & 1.9\ 10$^{17}$  & 10.6  & 77  & 0.46     &    & 4.7   & 1.3 & 325  & 149  & \cite{boord74}\\
  CdCr$_2$S$_4$    &     & 10.6  & 77  & 0.25     & 32 & 1.75  & 2.9 & 270  & 67   & \cite{jacobs74}\\
  InSb & 2.0\ 10$^{16}$  & 10.6  & 78  & 0.49     & 20 & 0.7   & 1.2 & 328  & 160  & \cite{tomasetta79}\\
  InSb & 5.0\ 10$^{13}$  & 10.6  & 35  & 1.7      & 23 & 0.04  & 0.4 & 827  & 1406 & \cite{tomasetta79}\\
  InAs & 6.0\ 10$^{17}$  & 5.4   & 300 & 0.58     & 25 & 2.66  & 1.5 & 224  & 130  & \cite{carlisle89}\\
  \hline
  InSb & 2.4\ 10$^{17}$  & 9.166 & 300 & 1.16     & 27 & 3.86   & 1.0 & 258  & 299 & This work\\
  \hline
  InSb & 2.4\ 10$^{17}$  & 9.166 & 300 & 2.1      & $>$50 & 3.86& 0.6 &      &     & double stage\\
  \hline
\end{tabular}
\caption{Comparison of published mid-infrared Faraday isolator performances.}\label{tab_Faraday_comp}
\end{table}

The present results could be improved further in several ways. The simplest one is to use commercially available high-quality wire grid polarizers with 98\% transmission and 1:500 extinction ratio. With such polarizers, a 27~dB isolation ratio is expected, and the insertion losses could be reduced to 1.1~dB.

The Faraday isolator figure of merit $F_d$ can also be increased using higher $B$-fields and thinner wafers. Similar simple permanent magnet configurations giving 1.7~T and 2.1~T fields have been patented~\cite{vigue} or reported~\cite{trenec,mukhin}. With 2.1~T, the wafer thickness can be reduced to 215~$\mu$m, resulting in a 94.7\% wafer transmission (92\% including the residual reflection losses after coating) for a single stage isolator, i.e. a 0.6~dB insertion loss assuming high quality (98\% transmission) polarizers.

Very high isolation ratios are usually achieved using a double stage Faraday rotator, with two wafers and three polarizers. In that case, it is crucial to lower polarizer and wafer losses, since with presently reported values the overall transmission of such a device would be only $0.9^3 \times 0.8^2 = 0.467$ (for example, in H$_2^+$ two-photon spectroscopy the gain in transition probability is reduced to a factor of 2.4).
Using a 2.1~T field and high quality polarizers, one can expect a 79\% isolator transmission (1~dB insertion loss) with an isolation ratio exceeding 50~dB for a double stage isolator.

\section{Conclusion}

We have reported the analysis of Faraday rotation in n-doped InSb wafers and the operation of an optical Faraday isolator with a 30~dB isolation ratio and 1.9~dB insertion loss at 9.166~$\mu$m. We have also discussed possible improvements and shown the feasibility of a room temperature double stage isolator which would ensure $>$50~dB isolation ratio with 1~dB insertion loss. Such a device significantly extends the range of QCL applications , especially in high-resolution spectroscopy.

The present results can be extended to the transparency range of InSb. The short wavelength limit is $\lambda \approx\ $7.5~$\mu$m
due to the $0.17$~eV gap energy. The long wavelength limit is $\approx$~30~$\mu$m due to the plasma frequency cut-off $\omega_p/n$. For shorter wavelengths in the 4-8~$\mu$m range, n-doped InAs could be used, but with higher absorption and reduced performances~\cite{boord74}.

\section*{Acknowledgments}
The authors thank the training students A. Besse, A. Orgogozo and Ph. Genevaux for their participation in the experiments, as well as P. Grech for wafer polishing. This work was supported by the CNRS PEPS ``m\'etrologie du futur'' action.


\begin{thebibliography}{99}

\bibitem{faist94}J. Faist, F. Capasso, D. L. Sivco, C. Sirtori, A. L. Hutchinson, and A. Y. Cho,``Quantum cascade laser'', Science {\bf 264}, 553-556 (1994).
\bibitem{capasso10}F. Capasso,``High-performance midinfrared quantum cascade lasers'', Optical Ingineering {\bf 49}, 111102, 1-9 (2010).
\bibitem{hugi10}A. Hugi, R. Maulini, and J. Faist, ``External cavity quantum cascade laser'', Semiconductor Science and Technology {\bf 25}, 083001, 1-14 (2010).
\bibitem{mansfield80}D. K. Mansfield, A. Semet, and L. C. Johnson, ``A lossless, passive isolator for optically pumped far-infrared lasers'', Appl. Phys. Lett. {\bf 37}, 688-690 (1980).
\bibitem{dennis68}J. H. Dennis,``A 10.6-micron four-port circulator using free carrier rotation in InSb'', IEEE J. Quantum Electron. {\bf 3}, 416 (1967).
\bibitem{boord74}W. T. Boord, Y.-H. Pao, F. W. Phelps, and P. C. Claspy, ``Far-infrared radiation isolator'', IEEE J. Quantum Electron. {\bf 10}, 273-279 (1974).
\bibitem{jacobs74}S. D. Jacobs, K. J. Teegarden, and R. K. Ahrenkiel,``Faraday rotation optical isolator for 10.6-$\mu$m radiation'', Appl. Opt.
    {\bf 13}, 2313-2316 (1974).
\bibitem{tomasetta79}L. R. Tomasetta, W. E. Bicknell, and D. H. Bates, ``100 W average power 10.6 $\mu$m isolator based on
    the interband Faraday effect in InSb'', IEEE J. Quantum Electron. {\bf 15}, 266-269 (1979).
\bibitem{aggarwal88}R. L. Aggarwal, R. F. Lucey, and D. P. Ryan-Howard, ``Faraday rotation in the 10 $\mu$m range in InSb at liquid-helium temperature'', Appl. Phys. Lett. {\bf 53}, 2656-2658 (1988).
\bibitem{carlisle89}C. B. Carlisle and D. E. Cooper, ``An optical isolator for mid-infrared diode lasers'', Opt. Commun. {\bf 74}, 207-210 (1989).
\bibitem{klein89}C. A. Klein and T. A. Dorschner,``Power handling capability of Faraday rotation isolators for CO$_2$ laser radar'', Appl. Opt. {\bf 28}, 904-914 (1989).
\bibitem{roth08}B. Roth, J. Koelemeij, S. Schiller, L. Hilico, J.-Ph. Karr, V. Korobov, and D. Bakalov, ``Precision spectroscopy of molecular hydrogen ions: towards frequency metrology of particle masses'', in {\it Precision Physics of Simple Atoms and Molecules}, Lecture Notes in Physics {\bf 745}, ed. S. Karshenboim (Springer, 2008), pp. 205-232.
\bibitem{korobov09}V. I. Korobov, L. Hilico, and J.-Ph. Karr, ``Relativistic corrections of m$\alpha^6$(m/M) order to the hyperfine
    structure of the H$_2^+$ molecular ion'', Phys. Rev. A {\bf 79}, 012501, 1-10 (2009).
\bibitem{bielsa07}F. Bielsa, A. Douillet, T. Valenzuela, J.-Ph. Karr, and L. Hilico, ``Narrow-line phase-locked quantum cascade laser in the 9.2~$\mu$m range'', Opt. Lett. {\bf 32}, 1641-1643 (2007).
\bibitem{bielsa08}F. Bielsa, K. Djerroud, A. Goncharov, A. Douillet, T. Valenzuela,  C. Daussy, L. Hilico, and A. Amy-Klein, ``HCOOH high-resolution spectroscopy in the 9.18 $\mu$m region'', J. Molec. Spectrosc. {\bf 247}, 41-46 (2008).
\bibitem{karr08a}J.-Ph. Karr, F. Bielsa, A. Douillet, J. Pedregosa Gutierrez, V. I. Korobov, and L. Hilico, ``Vibrational spectroscopy of H$_2^+$: Hyperfine structure of two-photon transitions'', Phys. Rev. A {\bf 77}, 063410, 1-10 (2008).
\bibitem{karr08b} J.-Ph. Karr, V. I. Korobov, and L. Hilico, ``Vibrational spectroscopy of H$_2^+$: precise evaluation of the Zeeman effect'', Phys. Rev. A {\bf 77}, 062507, 1-9 (2008).
\bibitem{Boer90}K. W. B\"oer, {\it A survey of semiconductor physics, 2nd edition}, (John Wiley \& Sons, 2002).
\bibitem{ioffe-web}http://www.ioffe.ru/SVA/NSM/Semicond/InSb/index.html
\bibitem{Zawadski}W. Zawadski, ``Electron transport phenomena in small-gap semiconductors'', Advances in Phys. {\bf 23}, 435-522 (1974).
\bibitem{vigue}J. Vigu\'e, G. Tr\'enec, O. Cugat, and W. Volondat, Patent n$^{\circ}$ WO 2008/031935 A1 30/03/2008.
\bibitem{trenec}G. Tr\'enec, W. Volondat, J. Vigu\'e, O. Cugnat, ''Permanent magnets for Faraday rotators inspired by the design of the magic sphere'', in preparation.
\bibitem{mukhin}I. Mukhin, A. Voitovich, O. Palashov, and A. Khazanov, ``2.1 Tesla permanent-magnet Faraday isolator for subkilowatt average power lasers'', Opt. Commun. {\bf 282}, 1969-1972 (2009).

\end{thebibliography}
\end{document}